\documentstyle[paspconf,11pt]{article}

\newcommand{\bc}{\begin{center}}
\newcommand{\ec}{\end{center}}
\newcommand{\bi}{\begin{itemize}}
\newcommand{\ei}{\end{itemize}}
\newcommand{\bd}{\begin{description}}
\newcommand{\ed}{\end{description}}
\def\hub{$H_0=100$ km s$^{-1}$ Mpc$^{-1}$, $q_0 = 0.5$}

\def\deg{\ifmmode $\setbox0=\hbox{$^{\circ}$}$^{\,\circ}
          \else    \setbox0=\hbox{$^{\circ}$}$^{\,\circ}$\fi\,}
\def\pdeg{\ifmmode $\setbox0=\hbox{$^{\circ}$}\rlap{\hskip.11\wd0 .}$^{\circ}
          \else \setbox0=\hbox{$^{\circ}$}\rlap{\hskip.11\wd0 .}$^{\circ}$\fi~}
\def\arcs{\ifmmode {^{\scriptscriptstyle\prime\prime}}
          \else $^{\scriptscriptstyle\prime\prime}$\fi~}
\def\arcm{\ifmmode {^{\scriptscriptstyle\prime}}
          \else $^{\scriptscriptstyle\prime}$\fi~}

\begin{document}

\bc
{\large \bf Status of VLBI Observations at 1\,mm Wavelength and Future Prospects}\\
\ec

\bc
S. Doeleman$^1$ and T.P. Krichbaum$^2$  \\

\vskip 0.25cm
\footnotesize{
$^1$ NEROC-MIT Haystack Observatory, Westford, MA., USA\\
$^2$ Max-Planck-Institut f\"ur Radioastronomie, Bonn, Germany\\
}
\ec

\vskip 0.5cm\leftline{\bf Introduction}
\vskip 0.25cm

One of the main motivations for high angular resolution imaging at wavelengths shorter than 
3\,mm comes from the need to image the innermost structures of AGN and their jets on scales 
as close as possible to the Schwarzschild radii of the central supermassive objects. 
At a wavelength of 1.3\,mm ($\sim 230$\,GHz), VLBI observations with transcontinental
baselines yield angular resolutions as small as 25 micro-arcseconds ($\mu$as). This corresponds
to a spatial resolution of 52 light days for 3C\,273 (z=0.158, \hub). Since most compact
radio sources are self-absorbed at longer wavelengths (or scatter broadened in the
case of Sgr\,A*), VLBI imaging at 1 or 2\,mm wavelength should offer a clear 
view to the nucleus, less affected by opacity effects. 

While VLBI imaging at 3\,mm with the CMVA (Coordinated Millimeter VLBI Array)
is now fairly standard (maps with dynamic range of a few hundred are obtained from experiments
involving up to 12 stations), VLBI observations at shorter wavelengths are still 
limited to single baseline detection experiments. With the foreseeable detection of
1\,mm fringes on short baselines also on both sides of the Atlantic, as a next step, the
combination of the European and American sub-arrays can be envisaged. With a 
global array operating near $\lambda=$1\,mm ($\nu=230$\,GHz) and utilizing
phased interferometers (at Plateau de Bure, Ovro, and Hat Creek) as sensitive elements, 
micro-arcsecond VLBI imaging should become possible within the next few years.

\vskip 0.5cm\leftline{\bf Summary of Previous Experiments}
\vskip 0.25cm

The first VLBI tests at the highest frequencies yet attempted ($1.3-1.4$\,mm)
were carried out in 1990, 1994, and 1995.
These tests were mainly technically driven and were performed to demonstrate the
feasibility of 1\,mm VLBI. A first experiment performed in 1990
yielded weak (SNR$=5$) fringes on 3C\,273 on the 845\,km baseline Ovro-Kitt Peak
(Padin et al., 1990). After this, two VLBI experiments were performed
in 1994 and 1995, using the 1150\,km baseline between the 30\,m MRT at 
Pico Veleta (Spain) and a single 15\,m antenna of the IRAM interferometer on Plateau de Bure 
(France). On this baseline fringes with a fringe spacing of $0.2-0.4$\,mas are obtained.
With system equivalent flux densities of SEFD$=2800$\,Jy for Pico Veleta and SEFD$=11500$\,Jy 
for Plateau de Bure, respectively, the single baseline detection threshold (7\,$\sigma$)
is 1.3\,Jy for incoherent averaging and 0.5\,Jy, if coherent averaging can be applied
(see below).

The first of these two experiments was made in December 1994.
This experiment was solely technically driven
and was performed to demonstrate the feasibility of 1\,mm VLBI using the
two IRAM instruments. After observations at 86\,GHz which were
used to determine the station clock offsets, the sources 3C\,273 
(${\rm S}_{\rm 215 GHz}= 13.5$\,Jy), 3C\,279 (${\rm S}_{\rm 215 GHz}= 10.5$\,Jy),
and 2145+067 (${\rm S}_{\rm 215 GHz}= 5.6$\,Jy) were observed and detected
with signal-to-noise ratios in the range of SNR=$7-10$ (Greve et al., 1995). 
1823+568 which had a flux of only 1.5\,Jy was not seen. 

A second observation took place in March 1995, using the same antennas
and observational setup (MK\,III mode A, 112\,MHz bandwidth). In this experiment which was
of longer duration, a sample of 8 bright AGN and the Galactic Center source
Sgr\,A* were observed (Krichbaum et al. 1997 \& 1998). From this small sample, only the 
faintest source, 4C\,39.25, which had a flux of ${\rm S}_{\rm 215 GHz}= 3.5$\,Jy, was not detected.
For the remaining 8 objects (3C\,273, 3C\,279, 1334-127, 3C\,345,
NRAO\,530, Sgr\,A*, 1749+096, 1921-293) clear fringes were found with 
signal-to-noise ratios in the range of SNR=$6-35$. 

In 1994 and 1995
3C\,273 and 3C\,279 were observed at similar interferometric hour angles (IHA).
This facilitates a comparison of their correlated flux densities between the
two epochs. Whereas for 3C\,279 the total flux density and the visibility amplitudes 
between both experiments were similar (${\rm S}_{\rm corr} = 2.4 - 2.8$\,Jy at IHA$=4$),
the correlated flux in 3C\,273 (at IHA$=2-3$) increased by a factor of two
from ${\rm S}_{\rm corr} = 0.5$\,Jy in December 1994, to ${\rm S}_{\rm corr} = 1.0$\,Jy 
in March 1995. On the other hand the total flux of 3C\,273 decreased from 
13.5\,Jy to 9.2\,Jy. This and the superluminal motion seen in 3C\,273 at longer
wavelengths, can be regarded as evidence for structural variations in the jet
of 3C\,273 also at 215\,GHz.

The highest correlated flux of about 4\,Jy
was seen in 3C\,279. This corresponds to a visibility (or compactness) of about 40\,\%.
For the other sources the visibilities were lower, ranging between 10--30\,\%.
At present it is unclear, if these lower visibilities are 
due to residual calibration uncertainties, or if they indicate 
angular resolution effects. All of the sources were observed in snapshot mode
for only a limited time range. The beating in the visibility amplitude,
which is caused by the mas-to sub-mas structure of the individual source
and which often is quite pronounced in the 3\,mm data, easily
could cause an underestimate of the correlated flux density and therefore
would represent only a lower limit to the compactness.

Recent 3\,mm maps of 3C\,273 (T. Krichbaum et al., this conference) show a one-sided
core jet structure with a compact core of $\sim 80$\,$\mu$as size and 
a brightness temperature ${\rm T_B}=2.1 \cdot 10^{11}$\,K close to the theoretically
expected inverse Compton limit (${\rm T_B} \sim 10^{12}$\,K). Using this measured brightness
temperature, the expected source size at 230\,GHz would be
$\theta = 10.5 \sqrt{\rm S_{Jy}}$\,$\mu$as. With a total flux density of
${\rm S}=10$\,Jy, the expected source size would be $33$\,$\mu$as.
This yields a visibility of $V=0.96$ or ${\rm S_{corr}} = 9.6$\,Jy at 700\,M$\lambda$ (Pico -- PdBure) and 
$V=3.9 \cdot 10^{-2}$ or ${\rm S_{corr}} = 0.4$\,Jy at 6000\,M$\lambda$ (transatlantic baselines).

In order to detect compact sources like 3C\,273 on the long transatlantic baselines, 
a detection sensitivity of $\simeq 0.4$\,Jy is needed. There might be
sources which are more compact, but they will be fainter. For a 1\,Jy source with
a size of $10$\,$\mu$as a correlated flux of $0.7-0.8$\,Jy could be expected, relieving
the sensitivity requirements by a factor of two. To reach the necessary sensitivity, 
future 1\,mm VLBI will require participation of antennas with large collecting 
areas (phased interferometers like Plateau de Bure, OVRO, BIMA, and the future 
MMA and ALMA), high observing bandwidths ($\Delta \nu  \geq 256$\,MHz), and the
possibility to correct for atmospheric phase fluctuations, which if uncorrected,
lead to too short coherence times (see Tahmoush \& Rogers, this conference).

\vskip 0.5cm\leftline{\bf A 1\,mm-VLBI Experiment in February 1999}
\vskip 0.25cm

Past 1mm-VLBI detections and scientific results are encouraging but
have, so far, been limited to single baselines.  For AGN studies at 1\,mm, 
imaging arrays are needed that include many more antennas.  Observations
of compact masers at the highest VLBI frequencies require relatively 
compact arrays with baselines less than $1G\lambda$.  A group of five 1mm
equipped mm-wave dishes in the SouthWest United States can potentially
deliver a scientifically useful 1mm-VLBI array.  This group includes :
the Berkeley-Illinois-Maryland Array (Redding, CA), the Owens Valley Radio
Observatory (Bishop, CA), the NRAO 12m (Kittpeak, AZ) and the Heinrich
Hertz Telescope (Mt. Graham, AZ).  

A 1mm-VLBI experiment was carried out with the above array plus the IRAM 30m
on Pico Veleta during the winter season in 1999.  Good weather is 
crucial for experiments at high frequencies especially since the smaller
antenna sizes typical of this array raise detection thresholds.  Recording
began at 0600 UT on 17 Feb and ended at 1800UT on Feb 19.  All sites 
recorded in MKIII compatible modes with VLBA sites (Kittpeak, OVRO) 
using 7 BBCs each 8MHz wide (56MHz BW), and all other sites using a full compliment
of 14 BBCs for a total of 112MHz bandwidth.  The center observing frequency 
was 230.5 GHz.

All sites other than the HHT routinely participate in CMVA 3mm-VLBI sessions
so special preparations there were required.  A MKIII VLBI electronics rack
and a tape recorder were shipped to Mt. Graham a month early and set up
for testing.  A H-maser was borrowed from the Harvard-SAO and shipped along
with the VLBI equipment.  Standard phase tests carried out at the HHT confirmed
that 1mm test tones injected into the receiver feed recorded properly on 
tape and could be recovered at the Haystack correlator.  As a further check,
the CO J=2-1 line at 230.5GHz was observed towards the cold core L1512, 
recorded using the VLBI system and its spectra generated by autocorrelating
the tape at Haystack.  Geodetic GPS measurements were made to determine the
HHT position to within 1 meter.
Personnel from the MPIfR
created the important software links from the VLBI field system to the HHT
pointing computer.

Target sources included the brightest compact AGN and placed emphasis on 
3C279 and 3C273B which were both at $\sim 10$Jy.  Historically, they have
been at higher flux density levels.  The compact source SgrA* was also 
included as it has a rising spectrum from 3 to 1mm.  

Bad weather at OVRO and BIMA made phasing the arrays difficult and we estimate
that during times of good mutual visibility of bright sources on the VLBI
array, these sites were unphased.  Baselines to Pico Veleta, while potentially
the most sensitive, were observed at very low elevations for antennas in the
US with correspondingly higher Tsys values.  The table shows the antenna 
sensitivities corresponding to the best times of mutual visibility on 3C273B and 
3C279.

\vskip 0.5cm
\begin{center}
\begin{tabular}{c|c|c|c} \hline\hline
Site & Diam & Tsys (K) & SEFD (Jy) \\ \hline
HHT & 10m & 375 & 22000\\
Kittpeak & 12m & 450 & 24000 \\
Pico Veleta & 30m & 300 & 2500 \\
BIMA & $9\times6$m  & 700 & $74000^*$ \\
OVRO & $4\times8$m & 1000 & $59000^*$ \\ \hline
\multicolumn{4}{l}{$^*$ Unphased} \\
\end{tabular}
\end{center}

\vskip 0.25cm\leftline{\bf Searching for Fringes}
\vskip 0.25cm

Tapes were shipped to the Haystack Correlator and fringes were searched
for to all sites.  Searches concentrated on the Kittpeak-HHT baseline
which observed 3C279 and 3C273B at optimal elevations and during periods
of good weather.  Station clocks determined using GPS receivers at each
site limited fringe searches in delay to a few micro seconds but delay
windows of up to $\pm28\mu sec$ were searched.  No sources were detected
using a combination of coherent and incoherent detection methods.

\vskip 0.5cm\leftline{\bf Sensitivity}
\vskip 0.25cm

An expression for the coherent detection threshold for a single
baseline can be written as :
\begin{equation}
D_c = \frac{7\sqrt{\mbox{SEFD}_1\;\mbox{SEFD}_2}}{L\;\sqrt{2B\tau_c}}
\end{equation}
\noindent where $L\sim0.5$ is the loss due to 1-bit sampling, B is the bandwidth,
and $\tau_c$ is the coherent integration time.  For the Kittpeak-HHT baseline,
the detection level for a coherence time of 10 seconds is 9.6 Jy.  This 
threshold can be lowered by averaging many coherent segments (Rogers, Doeleman \& 
Moran 1995) which is very useful in the high frequency regime where $\tau_c$ can
easily be less than 10 seconds.  This incoherent detection threshold ($D_i$) can
be expressed as $D_i\sim D_c\;N^{-0.25}$ where N is the number of coherent
segments averaged.  For scans of 6.5 minute length, the incoherent detection
threshold is lowered to $D_i=3.2$Jy.  The obvious question of why there were no
detections with a detection threshold well below the source flux densities leads
us to consider sources of loss in the VLBI systems.

Test tones traced through the receiver systems at both Kittpeak and HHT revealed
no more than a 20\% signal loss.  This loss alone would cause $D_i$ to increase by
a factor of 1.25.  A more severe loss of signal can come from decorrelation due to 
phase noise on the maser reference.  The maser coherence loss is $\exp(-\sigma^2/2)$
with $\sigma=2\pi\nu\sigma_y(\tau)\tau$ where $\sigma_y(\tau)$ is the Allan Standard
Variance at the coherence time $\tau$ and $\nu$ is the observing frequency.  
Investigation into the performance of the HHT H-maser shows that it may have had 
an Allan Variance in excess of 8e-14 for a 10 second coherence time causing a 
50\% loss in signal.  By comparison, the Kittpeak H-maser, with a variance of 2.4e-14
has a loss of only 4\%.  Combining these two sources of loss raises $D_i$ to 6Jy.
This leaves even the incoherent detection method with only a marginal chance of 
detecting the source on this baseline.  If we consider other possible factors 
such as source resolution on the $200M\lambda$ baseline or coherence times less
than 10 seconds, then the situation becomes even worse.  We conclude that the sum
of these losses combined with uncooperative weather to raise detection thresholds
on this baseline above flux densities of our brightest targets.

\vskip 0.5cm\leftline{\bf Future Experiments - 2mm}
\vskip 0.25cm

A compromise between the elevated detection thresholds at 230GHz and the need
to explore VLBI at higher frequencies may be to attempt 2mm-VLBI.  The sensitivity
advantages are clear.  The atmospheric opacity is much lower and coherence times
are longer than at 1mm.  System temperatures decrease while source flux densities
rise : at 150GHz the flux density of 3C273B is 13Jy (down from its historical 
level of 18Jy).  Effects of phase noise in the maser reference also decrease.
Using 2mm SEFDs of 17500Jy and 19500Jy for the HHT and Kittpeak respectively, we
find that $D_i$(10 sec)=3.3Jy with all the above losses accounted for.  For the 
spectral line case, if we assume a line width of 0.5km/s, then $D_c$(10 sec)=100Jy.
A number of SiO masers in evolved stars exceed this flux density in the 2mm range
and may be the best class of fringe finders at the higher frequencies.  Plans are
underway for a 2mm-VLBI test involving HHT-Kittpeak-Pico Veleta in early 2000.

\vskip 0.5cm\leftline{\bf Acknowledgements}
\vskip 0.25cm

The Feb 1999 observations required a great deal of preparation and hard work
from people at the following observatories: BIMA, Haystack, HHT, IRAM, MPIfR, NRAO,
and OVRO. Their efforts are crucial to the success of these high frequency experiments
and we are grateful for their assistance and dedication.

\vskip 0.5cm\leftline{\bf References}
\begin{verse}
Greve, A., Torres, M., Wink, J.E., et al., 1995, Astron. Astrophys. Lett., 299, L33.\\
Krichbaum, T.P., Graham, D.A., Greve, A., et al., 1997, Astron. Astrophys. Lett., 323, L17.\\
Krichbaum, T.P., Graham, D.A., Witzel, A., et al., 1998, Astron. Astrophys. Lett., 335, L106.\\
Padin, S., Woody, D.P., Hodges, M.W., et al., 1990, ApJ Let., 360,11.\\
Rogers, A.E.E., Doeleman, S.S. \& Moran, J.M., 1995, AJ, 109, 1391.\\
\end{verse}
\end{document}